\documentclass[aps,pre,showpacs,amssymb]{revtex4}
\usepackage{amsfonts}
\usepackage{graphicx}
\usepackage{bm}

\def\de{e}

\begin{document}

\title{Symmetry breaking induced by random fluctuations for
  Bose-Einstein condensates in a double-well trap}

\author{J. Garnier}
\affiliation{
Laboratoire de Statistique et Probabilit{\'e}s,
Universit{\'e} Paul Sabatier,
118 Route de Narbonne, 31062 Toulouse Cedex 4, France,\\
Tel. (33) 5 61 55 62 20, 
Fax. (33) 5 61 55 60 89, 
Email: garnier@cict.fr
}

\author{F. Kh. Abdullaev}

\affiliation{Physical-Technical Institute of the Uzbek Academy of Sciences,
         G. Mavlyanov str. 2-b, 700084, Tashkent, Uzbekistan}

\date{\today}

\begin{abstract}
This paper is devoted to the study of the dynamics of  two 
weakly-coupled Bose-Einstein condensates confined in a 
double-well trap and perturbed by  random external forces.
Energy diffusion due to random forcing allows the system to visit 
symmetry-breaking states when the number of atoms exceeds a 
threshold value.
The energy distribution evolves to a stationary distribution
which depends on the initial state of the condensate only 
through the total number of atoms. This loss of memory of the 
initial conditions allows a simple and complete
description of the stationary dynamics of the condensate
which randomly visits symmetric and symmetry-breaking states.
\end{abstract}

\maketitle

\section{Introduction}
The experimental achievements of Bose-Einstein condensates (BECs)
with various dilute trapped alkali-metal atom gases have opened 
new possibilities to investigate macroscopic quantum effects 
\cite{dalfovo}.
Indeed it is well known that phase coherence plays a crucial role 
in macroscopic quantum systems, and different systems have been 
addressed:
Josephson effect in superconductors \cite{Barone}, weakly linked superfluid 
He-B reservoirs \cite{Avenel},
two coupled BECs in a double well trap \cite{Smerzi},
and coupled condensates in different hyperfine levels \cite{Williams}.
In this paper we shall address the quantum coherent tunneling  between
two BECs trapped in a double-well potential,
and we shall focus our attention to the influence of random external 
fluctuations. 
In superconductors the main origin of external noise is
thermal or quantum  fluctuations.
Mathematically the problem can be reduced to the {\it pendulum} equation 
(sine-Gordon equation) with external fluctuations 
$\phi_{tt} + \sin(\phi) + \gamma \phi_{t} = f(t)$ \cite{Likharev}
and the energy diffusion of the system can be studied.
In particular Stratonovich has studied this problem in the strong 
damping case $\gamma \phi_{t} + \sin(\phi)=  f(t) 
$ \cite{Stratonovich}.

In BECs new types of problems appear due to the existence of the trap 
potential.
Coherent phase phenomena in coupled BECs are described by a system of
ordinary differential equations for the imbalance in atomic
population $z$ and relative phase $\phi$.
This system can be considered as the equation for the {\it nonrigid 
pendulum}. The pendulum case recovered if the imbalance is small 
$z^2 \ll 1$.
Nonrigidity leads to new phenomena, and we can distinguish 
two regimes in the atomic imbalance evolution.
The first regime is the macroscopic quantum tunneling (MQT), 
when the atomic 
population is periodically
exchanged between wells with $\left< z \right> =0$,
which is analogous to the Josephson effect. 
The second regime is the macroscopic quantum localization (MQL)
phenomena, when the mean field nonlinearity excesses some 
critical value,
the mean value of the imbalance becomes nonzero 
$\left< z \right> \neq 0$, 
which corresponds to the localization of the  atomic population 
in one of the wells in the form of small-amplitude
oscillations around the bottom of the well.

It is therefore interesting to investigate the influence of 
fluctuations on the evolution of the system in these states.
In particular we would like to pay a particular attention to the
possible switching between these states.
This problem has a general character and it is important for the theory 
of nonlinear directional couplers
in nonlinear optics \cite{ADK,KA} and nonlinear dimers 
\cite{Scott}.
Periodic variations of the parameters of the trap have already
been considered, and resonant phenomena  have been exhibited for the 
weakly-coupled BEC \cite{abdullaev00a} and for the strongly overlapped BEC
\cite{elyutin}.
The influence of a periodic time-varying atomic scattering
length was addressed in \cite{abdullaev00b}.
Dynamical tunneling between regions of regular motion 
was shown to be involved by strong periodic modulation of the tunnel
coupling between the two modes \cite{salmond}.
Finally macroscopic quantum chaos driven by a time-periodic trap asymmetry
was predicted in \cite{hai}.
In this work we shall study the influence of random fluctuations
on the BEC dynamics trapped in a double-well trap.
We shall address the case of fluctuating zero point 
energies so that the energy difference $\Delta E(t)$
is a zero-mean random process. Such variations are induced by
small oscillations in the barrier laser position
\cite{Tsukada,Giovanazzi}.
Indeed, the barrier is generated by a Gaussian laser sheet 
focused near the center of the harmonic trap 
and small oscillations of the barrier-laser position introduce
the fluctuations of the zero-point energies.

\section{Bose-Einstein condensates in a double-well trap}

The problem of BECs
in a double-well time-dependent trap is described by the 
Gross-Pitaevskii (GP) equation
\begin{equation}\label{gp}
i\hbar\Psi_{t} =-\frac{\hbar^2}{2m}\Delta\Psi + V_{tr}(r,t)\Psi
+ g|\Psi|^{2}\Psi,
\end{equation}
where $V_{tr}$ is the double-well potential, $g = 4\pi\hbar^{2}a_{s}/m$,
and $a_{s}$ is the atomic scattering length.
For weakly overlapped condensates (high barrier or well separated wells)
we can use the two-modes decomposition
\begin{equation}
\Psi = \phi_{1}(t)\Phi_{1}(r) + \psi_{2}(t)\Phi_{2}(r),
\end{equation}
where $\psi_{1,2}$ are complex time-dependent amplitudes of condensates in wells and 
$\Phi_{1,2}$ are approximate ground state solutions of 
GP equation in first and second wells respectively.
Substituting this solution into (\ref{gp}), multiplying the equation
by $\overline{\Phi_{1,2}}$  
and integrating over the spatial variable we obtain the system of 
equations for the two modes $\psi_{1}(t),\psi_{2}(t)$
\cite{raghavan,marino}
\begin{eqnarray}
&&  i \hbar \frac{\partial \psi_1}{\partial t}
= [E_1(t)+ \alpha_1 |\psi_1|^2 ] \psi_1 - K \psi_2 ,\\
&& i \hbar \frac{\partial \psi_2}{\partial t}
= [E_2(t)+ \alpha_2 |\psi_2|^2 ] \psi_2 - K \psi_1 ,
\end{eqnarray}
where $E_{1,2}$ are the zero point energies, 
$K$ is the coupling constant between the two modes,
and $\alpha_{1,2}$ are the mean field nonlinearities for both modes.
Here we take into account a small oscillation in the laser-barrier
position so that $\Delta E = E_{1} - E_{2}$ is 
time-varying \cite{smerzi00}.
We write $\psi_j = \sqrt{N_j} \exp(i \theta_j)$.
$N_T=N_1+N_2$ is the (constant) total number of atoms.
The fractional population imbalance 
\begin{equation}
  z(t) = \frac{N_1(t) - N_2(t)}{N_T} 
\end{equation}
and the relative phase
\begin{equation}
  \phi(t) = \theta_2(t)-\theta_1(t)
\end{equation}
satisfy 
\begin{eqnarray}
\label{eq:sys1a}
&&  z_t = - \sqrt{1-z^2} \sin(\phi) -\eta \phi_t ,\\
\label{eq:sys1b}
&& \phi_t = -\Delta E(t) + \Lambda z + \frac{z}{\sqrt{1-z^2}}
\cos(\phi) ,
\end{eqnarray}
where we have rescaled to a dimensionless time $t 2 K / \hbar
\rightarrow t$
and we have  introduced new variables
\begin{eqnarray}
  &&\Delta E(t) =  \frac{E_1(t)-E_2(t)}{2K} +
  \frac{\alpha_1-\alpha_2}{4K} N_T ,\\
&&\Lambda = \frac{\alpha N_T}{2K} ,\  \ \ \
 \alpha= \frac{\alpha_1+\alpha_2}{2}.
\end{eqnarray}
Note that  we have included the damping term $-\eta \phi_t$ 
in Eq.~(\ref{eq:sys1a})
that takes into account a non-coherent dissipative current of
normal-state atoms, proportional to the chemical potential difference 
\cite{abdullaev00a}.

\section{Hamiltonian structure of the unperturbed system}
The Hamiltonian of the unperturbed system ($\Delta E=0$, $\eta=0$)
is
\begin{equation}
\label{unpertham}
  E = \frac{\Lambda z^2}{2} - \sqrt{1-z^2} \cos(\phi) .
\end{equation}
$E$ can take any value between $E_{min} = -1$ 
and $E_{max}  = \frac{1}{2} (\Lambda + \frac{1}{\Lambda}) $.
It is an integral of motion.
The orbits of the motion are closed, 
corresponding to periodic oscillations.
In order to explicit the 
periodic structure of the variables $z$ and $\phi$,
we introduce the action-angle variables.
The orbits are completely 
determined by the  energy $E$ imposed by the initial
conditions.
The period is denoted by ${\cal T}(E) $,
the motion is described by
\begin{eqnarray}
  &&z(t) = {\cal Z}(E,\theta(t)) ,\\
  &&\cos( \phi(t)) = {\cal C}(E,\theta(t)) ,\\
  &&\sin( \phi(t)) = {\cal S}(E,\theta(t)) ,
\end{eqnarray}
where ${\cal Z}$, $ {\cal C}$, and  ${\cal S}$
are smooth functions, periodic with respect to $\theta$ with period $2
\pi$, the angle satisfies 
\begin{equation}
  \frac{d \theta}{dt}  = \frac{2 \pi}{{\cal  T}(E) } ,
\end{equation}
and the action is defined by
\begin{equation}
  {\cal I}(E) = \frac{1}{2 \pi} \int_{-1}^E {\cal T}(x) dx .
\end{equation}

The analysis is standard but quite cumbersome, so we only list the
final results that can be written in terms of elliptic functions.
We need to introduce a series of parameters.
\begin{eqnarray}
  && \kappa(E) = \frac{\Lambda E-1}{2} , \ \ \ \ 
  \zeta(E) = \frac{\Lambda^2-1}{4} - \kappa(E) \\
  && M(E) = \frac{1}{2} \left( 1 + \frac{\kappa}{\sqrt{\zeta}} \right) ,
  \ \ \ \ 
  z_2(E) = \frac{2}{\Lambda} \sqrt{\kappa+ \sqrt{\zeta}}.
\end{eqnarray}
Three cases can be distinguished.

{\it 1.} If $E \in [-1,1)$, then
$M<1$, the period is 
\begin{equation}
  {\cal T}(E) = \frac{8 \sqrt{M}}{\Lambda z_2} K(M),
\end{equation}
where $K$ is the complete elliptic function \cite{abra},
\begin{eqnarray}
  && {\cal Z}(E,\theta) = z_2 {\rm cn} \left( \frac{2 K(M) \theta}{\pi} ,
  M \right) ,\\
  && {\cal C}(E,\theta) = \frac{\Lambda {\cal Z}^2
    -2E}{2\sqrt{1-{\cal Z}^2}} ,\\
  && {\cal S}(E,\theta) = \frac{\Lambda z_2^2}{2 \sqrt{M}
    \sqrt{1-{\cal Z}^2} } {\rm sn} \ {\rm dn} \left(  \frac{2 K(M)
      \theta}{\pi} ,  M \right) .
\end{eqnarray}
cn, sn, and dn are tabulated Jacobian functions \cite{abra}.
Note that ${\cal Z}$ is an even function with respect to $\theta$.
These solutions preserve the $z$-symmetry in the sense that
they correspond to eigenfunctions of the Gross-Pitaevskii
equation that are either odd or even functions.
The stationary ground state has energy $-1$ and it is given by $z \equiv 0$ and
$\phi\equiv 0$.
For $E \in (-1,1)$, we have $\left< z \right>=0$ 
and the atomic population is periodically exchanged between the two modes. 
This is the MQT regime.

{\it 2.} If $E \in (1,E_{max})$, then
$M>1$, the period is 
\begin{equation}
  {\cal T}(E) = \frac{4}{\Lambda z_2} K(\frac{1}{M}) .
\end{equation}
The three periodic functions ${\cal Z}$, ${\cal C}$, and ${\cal S}$ are
given by
\begin{eqnarray}
  && {\cal Z}(E,\theta) = \pm 
   z_2 {\rm dn} \left( K(\frac{1}{M}) \frac{\theta}{\pi} ,
  \frac{1}{M} \right) ,\\
  && {\cal C}(E,\theta) = \frac{\Lambda {\cal Z}^2
    -2E}{2\sqrt{1-{\cal Z}^2}} ,\\
 && {\cal S}(E,\theta) = \frac{\Lambda z_2^2}{2 M
    \sqrt{1-{\cal Z}^2} } {\rm sn} \ {\rm cn} \left( K(\frac{1}{M})
 \frac{ \theta}{\pi} ,  \frac{1}{M} \right) .
\end{eqnarray}
Note that ${\cal Z}$ is either always positive, or always
negative-valued.
These solutions correspond to eigenfunctions
that break the $z$-symmetry.
In particular there exist two stationary solutions
with  the maximal energy $E_{max}$ that are
 $z\equiv \pm \sqrt{1-1/\Lambda^2}$,
$\phi\equiv \pi$.
Their existences result from the nonlinear interatomic interaction
and it is possible only when the number of atoms
is large enough so that nonlinear coefficient $\Lambda >1$.
If $E \in (1,E_{max})$ the population imbalance periodically oscillates
around a non-zero average value $\left< z \right> \neq 0$.
This is the MQL regime. 

{\it 3.} If $E =1$, then
$M=1$, the motion is given, in this special case, by the
non-oscillatory hyperbolic secant
\begin{equation}
  z(t) =     z_2 {\rm sech} \left(  \sqrt{\Lambda - 1} (t+t_0) \right) ,
\end{equation}
with $z_2= 2 \sqrt{\Lambda - 1}/\Lambda$ and
$t_0 = (1/\sqrt{\Lambda-1}) {\rm argch}(z_2/z_0)$.

\section{Effective dynamics}

\subsection{Damping}
\begin{figure}
\begin{center}
\begin{tabular}{cc}
{\bf a)}
\includegraphics[width=7.cm]{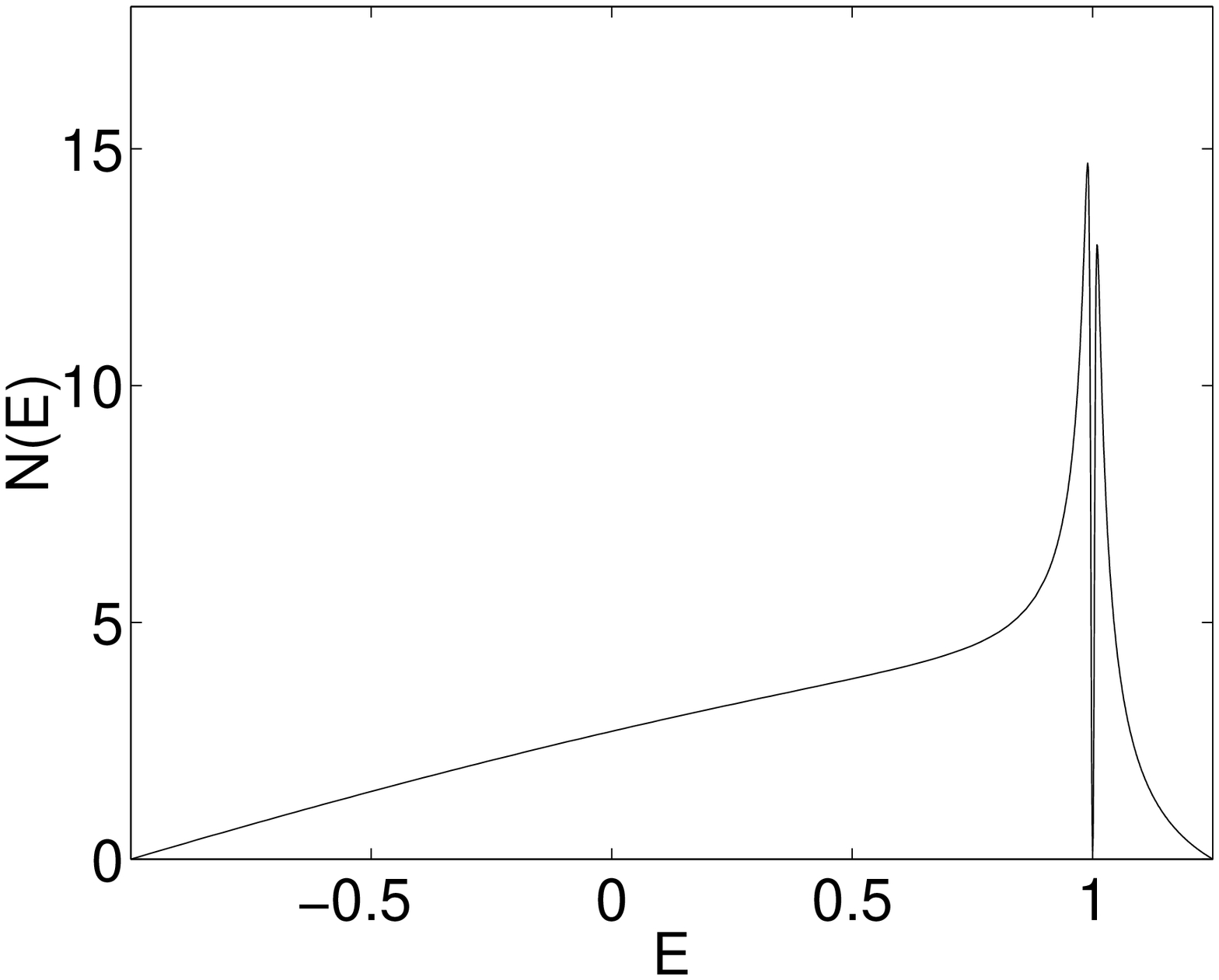}
&
{\bf b)}
\includegraphics[width=7.cm]{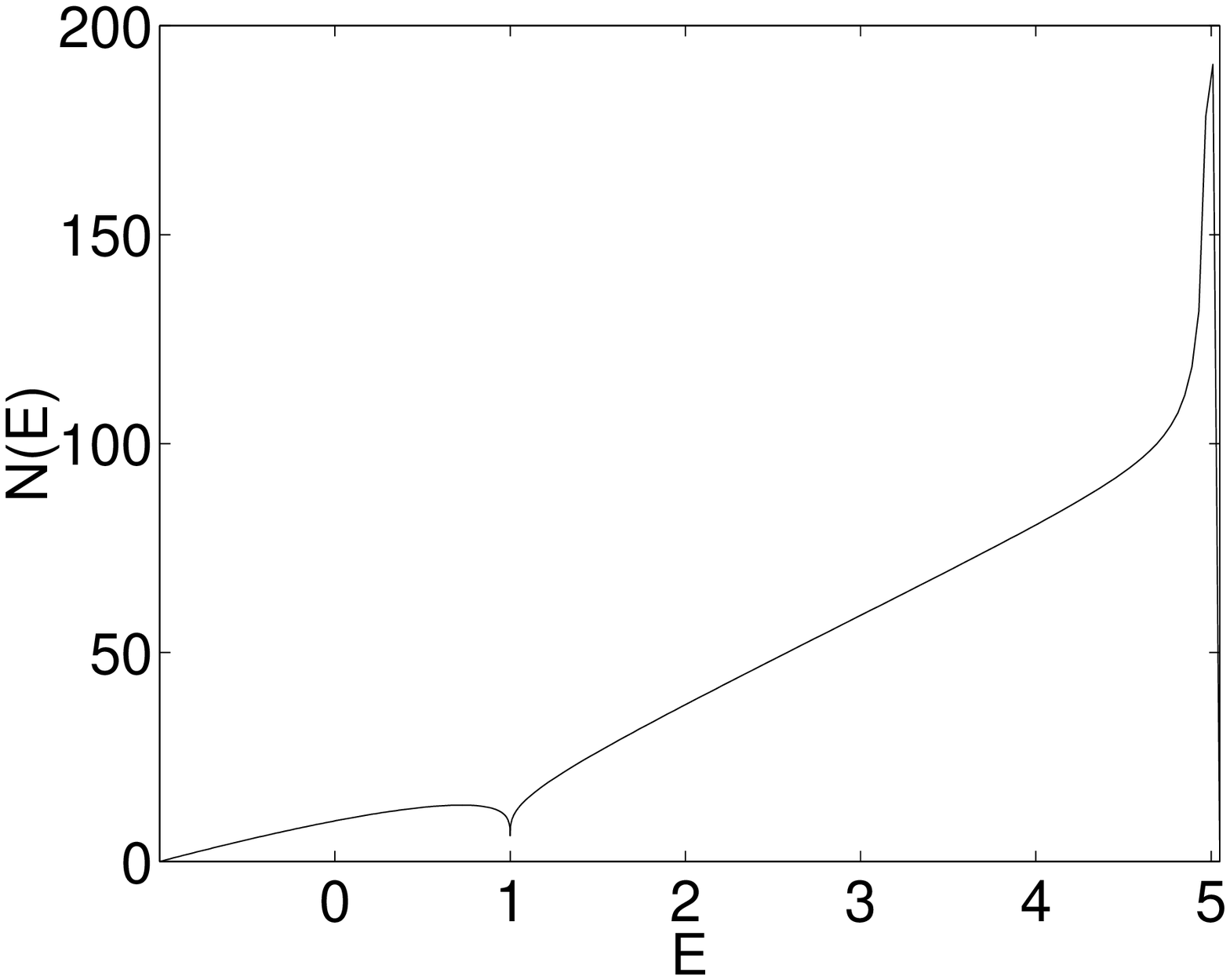}
\end{tabular}
\end{center}
\caption{Damping term $E \mapsto {\cal N}(E)$ for $\Lambda =2$ (picture a)
and $\Lambda =10$ (picture b).
\label{figdamp1}
}
\end{figure}

In presence of a small damping an adiabatic approach is possible
which yields that the energy decays as
\begin{equation}
\label{eq:effdamp}
  \frac{dE}{dt} = - \eta {\cal N}(E) ,
\end{equation}
where
\begin{equation}
  {\cal N}(E)  = \frac{1}{2\pi } \int_0^{2 \pi} 
\left(\Lambda {\cal Z} +\frac{\cal Z}{\sqrt{1-{\cal Z}^2}}
{\cal C}  \right)^2 (E,\theta) d\theta .
\end{equation}
${\cal N}(E)$ vanishes at $E=-1$, $E=1$, and $E=E_{max}$.
It takes large values just above and below $E=\Lambda /2$.
In Figure \ref{figdamp1} we plot the function $E \mapsto {\cal N}(E)$
for two different values of $\Lambda$.
Expansions of ${\cal N}$ are also presented in the Appendix.

\begin{figure}
\begin{center}
\begin{tabular}{cc}
{\bf a)}
\includegraphics[width=7.cm]{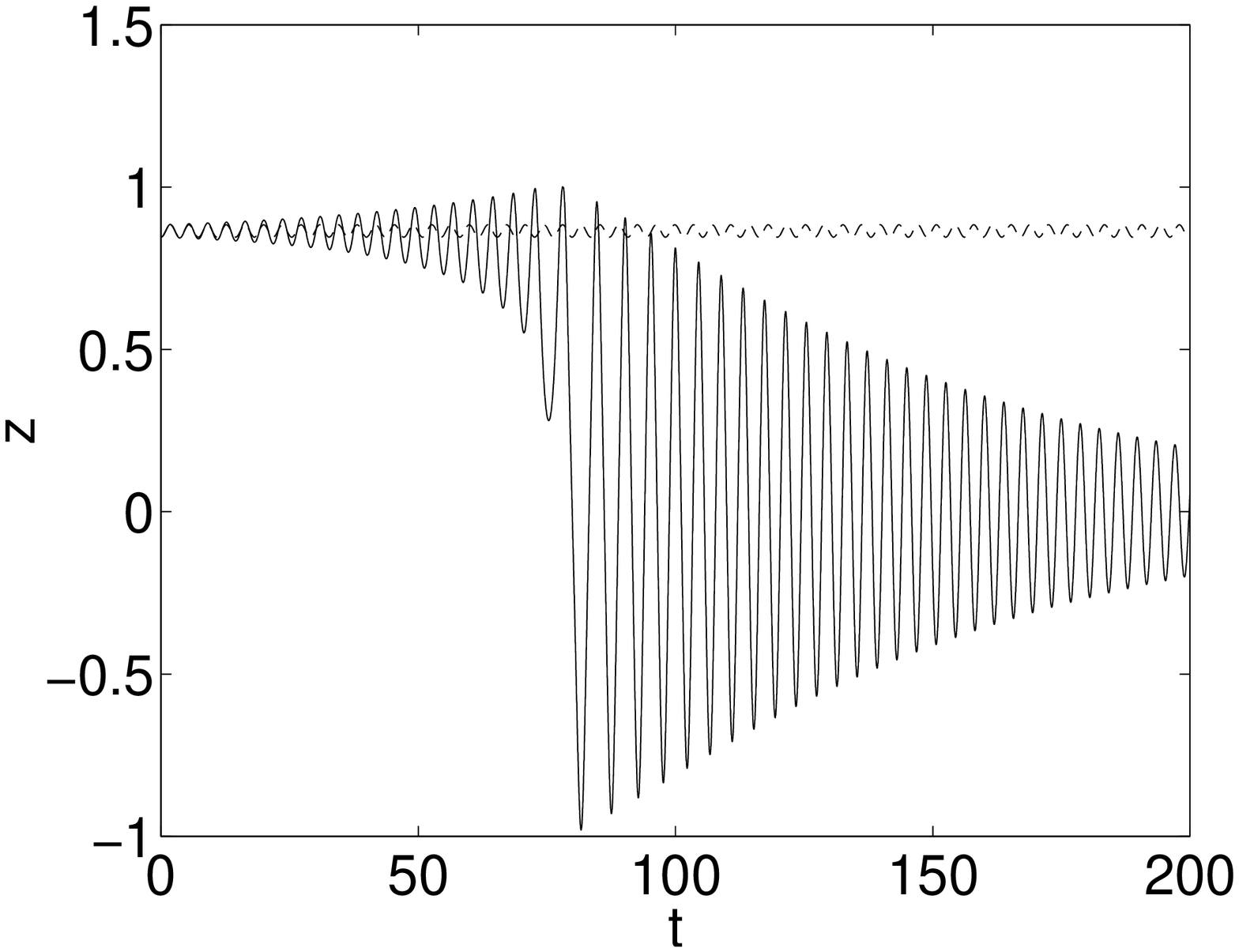}
&
{\bf b)}
\includegraphics[width=7.cm]{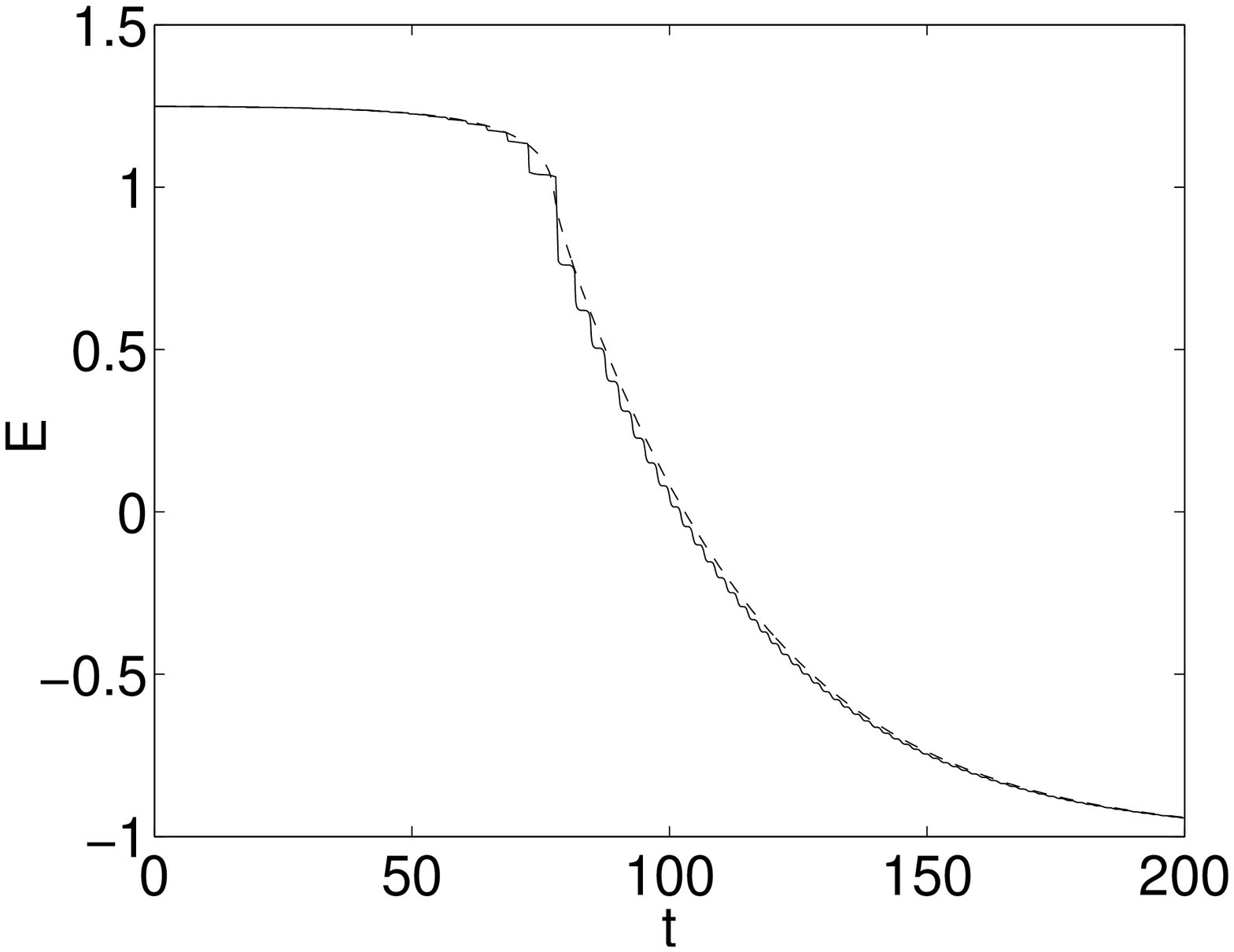}
\end{tabular}
\end{center}
\caption{Picture a:
Evolution of the population imbalance $z$ in absence (dashed line)
and in presence (solid line) of a small damping $\eta=10^{-2}$.
Here $\Lambda=2$ and the initial state 
is $z_0=\sqrt{1-1/\Lambda^{2}}-0.02 \simeq 0.846$ and $\phi_0=\pi$.
Picture b: Evolution of the energy $E(t)= 
{\Lambda z^2(t)}/{2} - \sqrt{1-z^2(t)} \cos(\phi(t))$
according to the numerical resolution of the system 
(\ref{eq:sys1a}-\ref{eq:sys1b}) (solid line)
and according to the effective equation (\ref{eq:effdamp}) (dashed line).
\label{figdamp2}
}
\end{figure}

Damping is not able by itself to break the symmetry as it involves an
energy decay, while symmetry breaking occurs when the energy
goes from a value below the separatrix $1$ to a value above
the separatrix.
If the system starts from one of the two breaking states with the
energy $E_{max}$ 
then it stays exactly at this state as ${\cal N}(E_{max})=0$.
However, a linear stability analysis shows that these states are not stable.
More exactly, if the initial state has energy  $E_{max}-e$  with $e \ll 1$,
then it stays close to this state during a time of the 
order of $\sim [\Lambda (\Lambda^2-1) e]^{-1}$ because
${\cal N}(E_{max}-e) \simeq \Lambda (\Lambda^2-1) e$;
after this time it quits 
the breaking state and converges to the ground state $E_{min}$.
This behavior is described in Figure \ref{figdamp2}.
Note the remarkable agreement between the results
from numerical simulations of the system of equations
(\ref{eq:sys1a}-\ref{eq:sys1b}) with $\Delta E=0$
and the effective solution given by Eq.~(\ref{eq:effdamp}).

\subsection{Random fluctuations}
Let us consider a random perturbation of the system
that can be written in the form
\begin{eqnarray}
  && z_t = - \partial_\phi H ,\\
  && \phi_t = \partial_z H,
\end{eqnarray}
where $H = H_0(z,\phi) + m(t) V(z,\cos(\phi),\sin(\phi) )$, $H_0$
is the unperturbed Hamiltonian (\ref{unpertham}), and $m(t) V$ is the
time-dependent perturbation.
$m$ is assumed to be a stationary, zero-mean, random process.
We think in particular at a time-dependent $\Delta E(t)$
in Eq.~(\ref{eq:sys1a}) so that $m(t) = \Delta E(t)$ and $V=-z$.

In  presence of perturbations, 
the motion of $(z,\phi)$ is not purely oscillatory,
because the energy and the action are slowly varying in time.
We adopt the action-angle formalism, because it allows us to 
separate the fast scale of the locally periodic motion and the slow
scale of the evolution of the action. The motion is governed 
by the system
\begin{eqnarray}
  && I_t = - m(t) h_\theta(I,\theta) ,\\
  && \theta_t = \omega(I)
+ m(t) h_I(I,\theta) ,
\end{eqnarray}
where $\omega(I) =  \frac{2 \pi}{{\cal T}\circ {\cal E}(I)}$,
 $I\mapsto {\cal E}(I)$ is the inverse function of $E
\mapsto {\cal I}(E)$,
 and $h(I,\theta)= V( {\cal Z}(I,\theta) ,
{\cal C}(I,\theta) ,{\cal S}(I,\theta) )$.
Using standard diffusion-approximation theory \cite{PSV},
we get that the action $I$ behaves like a diffusion Markov 
process with the infinitesimal generator
\begin{equation}
  {\cal L}_I =\frac{1}{2} \frac{\partial}{\partial I} 
\left[ A(I) \frac{\partial}{\partial  I } \right] .
\end{equation}
The diffusion coefficient is
\begin{equation}
  A(I) = \frac{1}{\pi} \int_0^{2 \pi} \int_0^\infty
h_\theta(I,\theta) h_\theta(I, \theta+ \omega(I) t ) 
\left< m(0) m(t) \right> dt d \theta ,
\end{equation}
where the brackets stand for a statistical averaging.
This means in particular that the probability density function of
${I}(t)$ satisfies the Fokker-Planck equation 
$\partial_t p = {\cal L}_I^* p$,
$p(t=0,{I})  = \delta({I}-I_0)$, where $I_0$
is the initial action at time $0$ and ${\cal L}_I^*$
is the adjoint operator of ${\cal L}_I$, 
which is equal to ${\cal L}_I$ in our configuration 
as ${\cal L}_I$ is self-adjoint.

\subsection{Energy diffusion}
\begin{figure}
\begin{center}
\begin{tabular}{cc}
{\bf a)}
\includegraphics[width=7.cm]{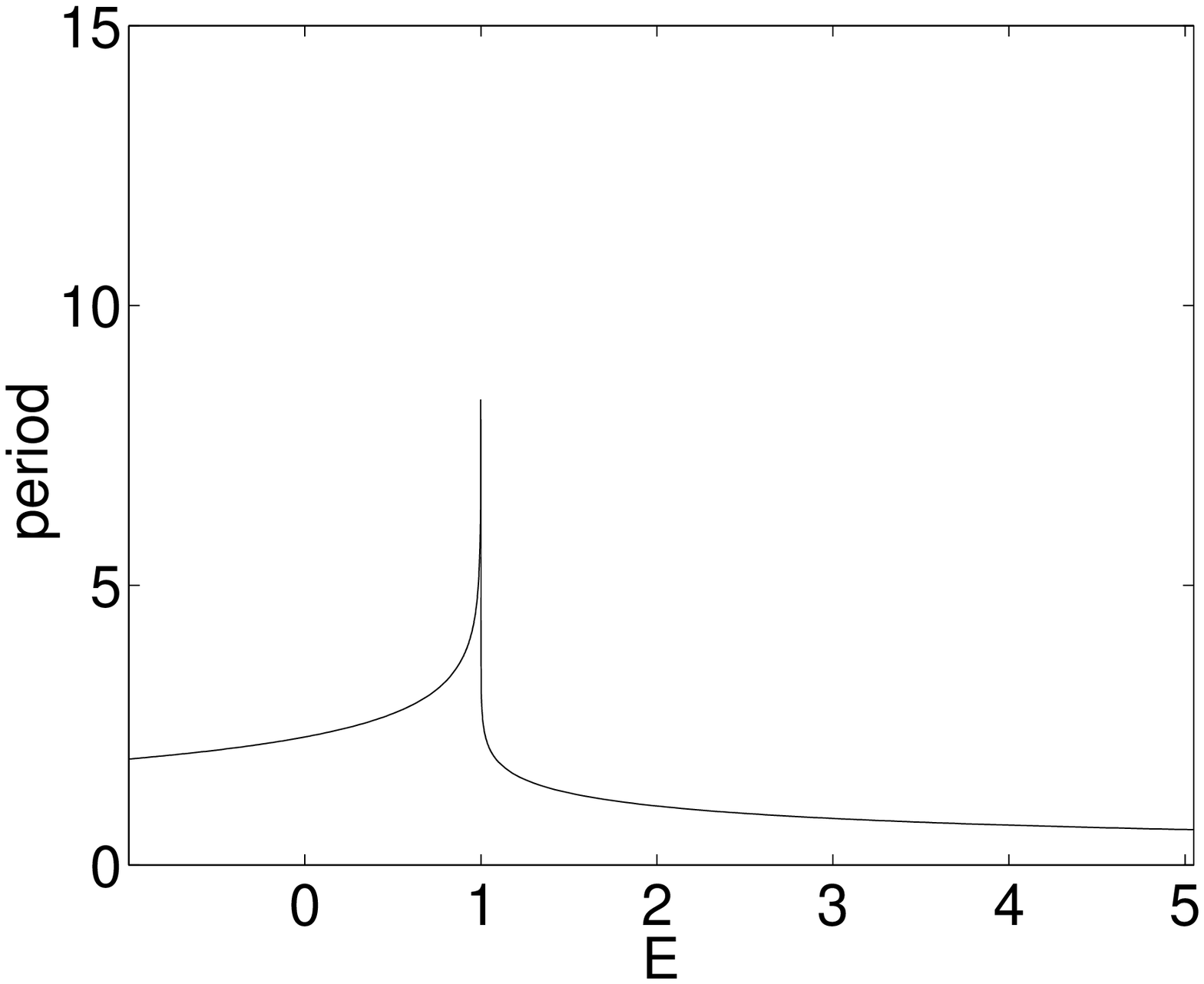}
& {\bf b)}
\includegraphics[width=7.cm]{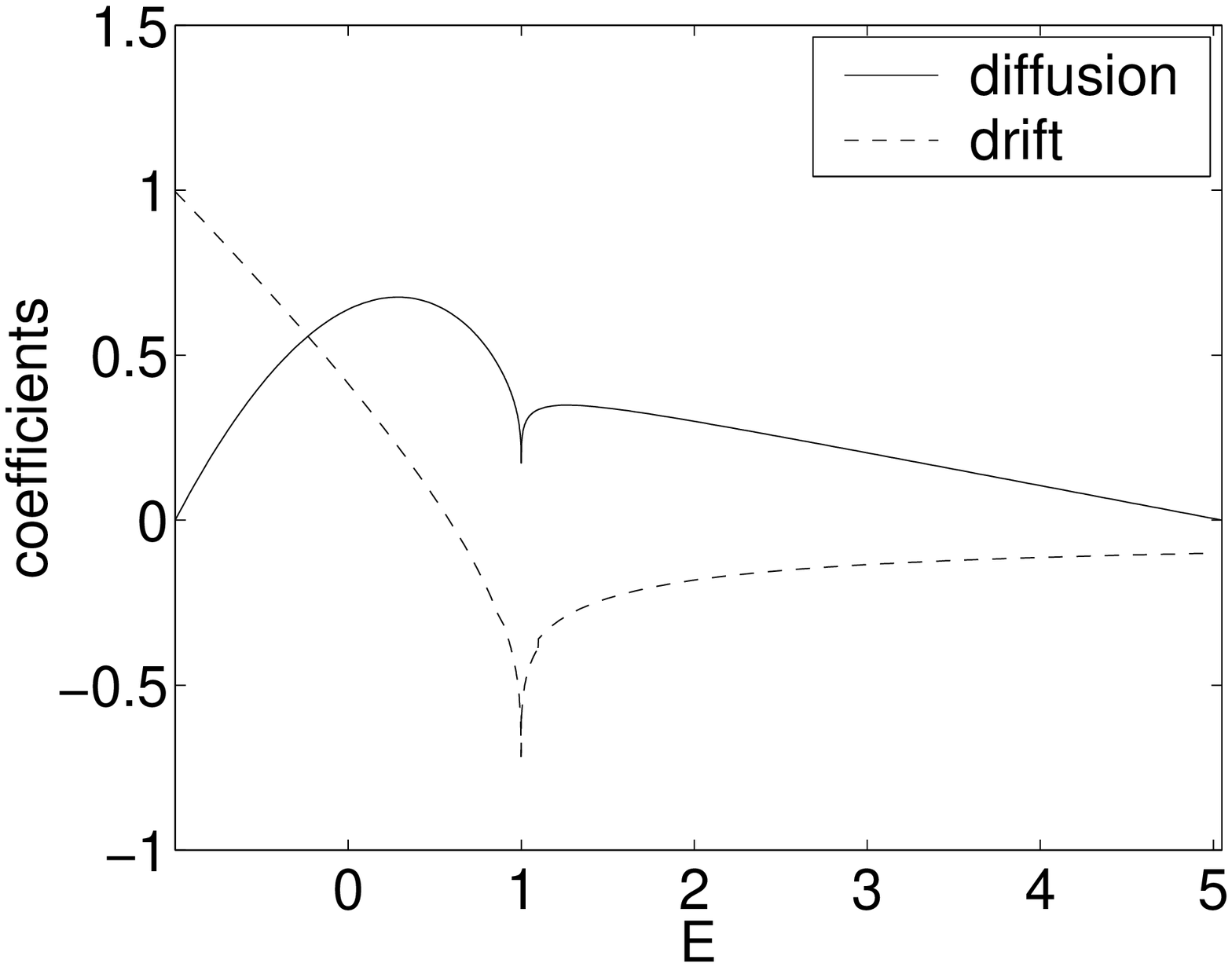}
\end{tabular}
\end{center}
\caption{Period ${\cal T}(E)$, drift and diffusion  coefficients
for $\Lambda =10$.
\label{figcoeff}
}
\end{figure}

The results of the two previous subsections can be combined 
to address the case of system (\ref{eq:sys1a}-\ref{eq:sys1b})
with a white noise model for $\Delta E$ and a damping  $\eta >0$.
We get that the energy $E$ of the system 
is a diffusion process with  the infinitesimal generator
\begin{equation}
  {\cal L}_E = \frac{2 \pi^2 \alpha}{{\cal T}(E)}
\frac{\partial}{\partial E}\left[ \frac{{\cal B}(E)}{{\cal T}(E) }
\frac{\partial}{\partial  E} \right] 
- \eta {\cal N}(E) \frac{\partial}{\partial E} ,
\end{equation}
where
\begin{eqnarray}
  {\cal B}(E) =  \frac{1}{2\pi} \int_0^{2 \pi} 
{\cal Z}_\theta(E,\theta)^2 d \theta ,\\
\alpha= \int_0^\infty \left< \Delta E(0) \Delta E(t) \right> dt .
\end{eqnarray}
The infinitesimal generator is not self-adjoint.
The drift of the energy is $d(E) = \alpha 2 \pi^2 \partial_E[
{\cal B}(E)/{\cal T}(E)] / {\cal T}(E) - \eta {\cal N}(E)$
and the diffusion coefficient  is $\sigma(E) = \alpha
2 \pi^2 {\cal B}(E)/{\cal T}^2(E)$.
They are plotted in Figure \ref{figcoeff} in  the case $\eta =0$.
This result allows us to compute all relevant quantities,
in particular the stationary energy distribution.
Indeed the diffusion process $E(t)$ is ergodic, and the statistical
distribution of the energy becomes independent of the initial
state for large time $t$. It converges to a stationary
distribution that can be computed explicitly as the 
solution of the elliptic equation ${\cal L}_E^* p=0$.

\section{Energy distribution}
\label{sec:energy}%
\begin{figure}
\begin{center}
\begin{tabular}{cc}
{\bf a)}
\includegraphics[width=7.cm]{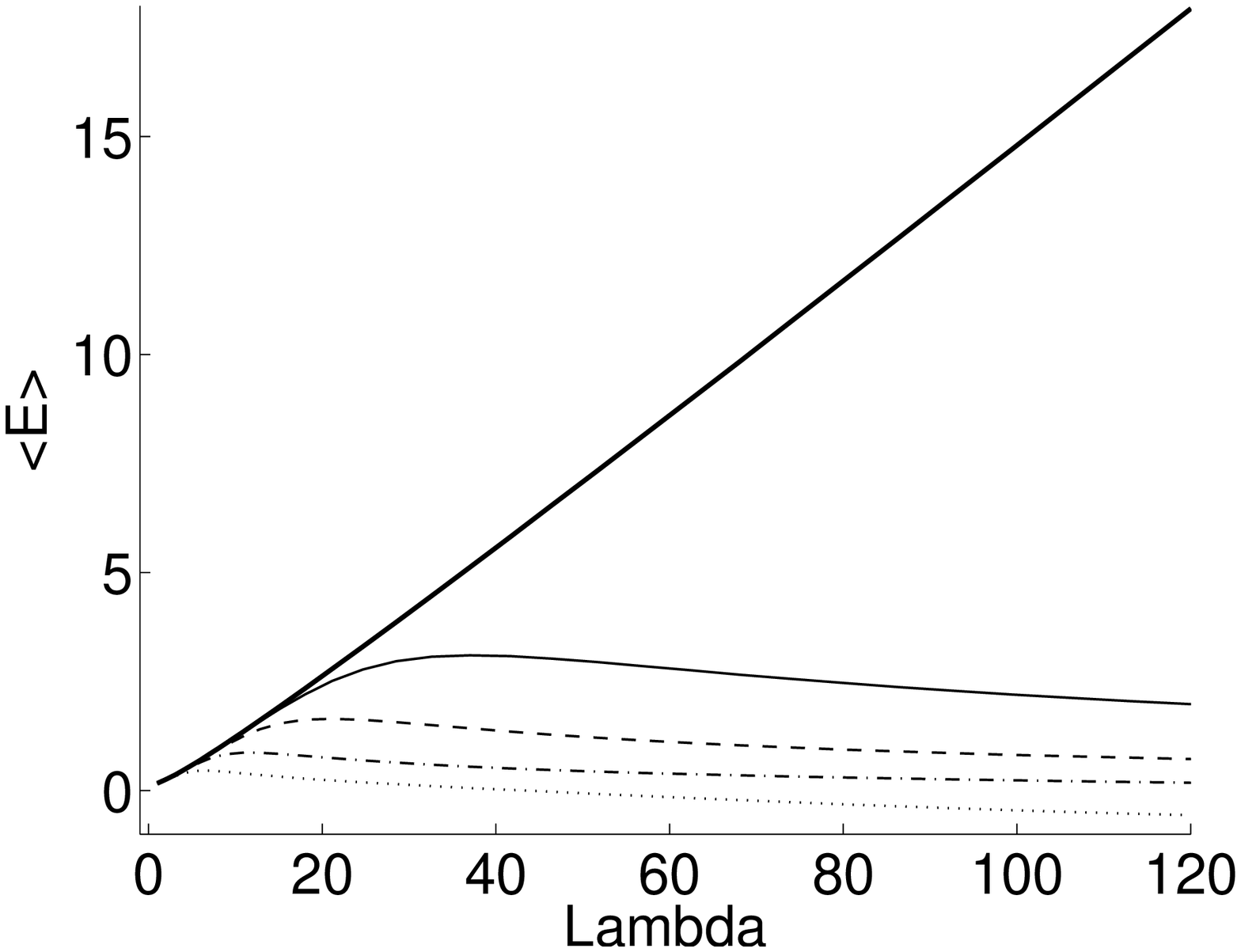}
&
{\bf b)}
\includegraphics[width=7.cm]{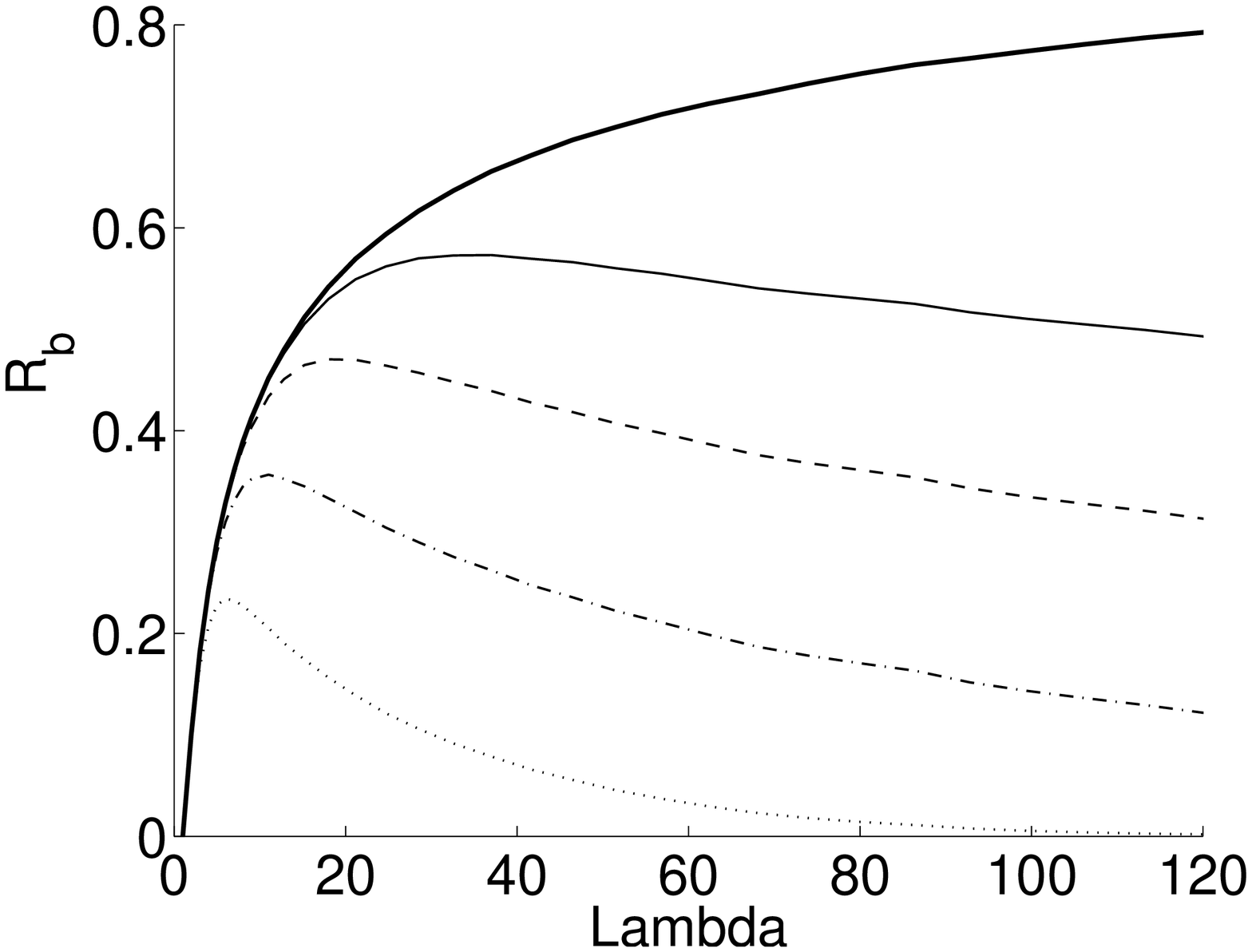}
\end{tabular}
\end{center}
\caption{Picture a:
Average energy as a function of $\Lambda$ for 
different damping rates.
Picture b:
Proportion of time spent by the system 
in breaking states $R_b$ as a function of $\Lambda$ for 
different damping rates: $\eta/\alpha=0$ (thick solid line),
$\eta/\alpha=10^{-5}$ (thin solid line),
$\eta/\alpha=10^{-4}$ (dashed line),
$\eta/\alpha=10^{-3}$ (dash-dotted line),
$\eta/\alpha=10^{-2}$ (dotted line).
\label{figrb}
}
\end{figure}

The stationary energy distribution for the system
is determined by the resolution of the equation ${\cal L}_E^* p=0$
that is 
\begin{equation}
\label{eq:statE}
  p(E)  = C {\cal T}(E) \exp \left( - \frac{\eta}{\alpha}
 \int_{-1}^E \frac{
{\cal T}^2 {\cal N}}{2 \pi^2 {\cal B} }(x) dx \right) ,
\end{equation}
where the normalization constant $C$ is
chosen so that $\int_{-1}^{E_{max}} p(E) dE =1$.
This distribution depends only on $\Lambda$ and the ratio 
$\eta/\alpha$.
A first important quantity is the average energy
\begin{equation}
\left< E \right> = \int_{-1}^{E_{max}} E p(E) dE.
\end{equation}
It is plotted in Figure \ref{figrb}a as a function of $\Lambda$
for different damping rates, which shows in particular that
the average energy grows linearly with $\Lambda$ when $\eta=0$,
according to $E \simeq \Lambda /6$,
but possesses a maximum if $\eta >0$.
Another important quantity is
the proportion of time  spent by the system 
in breaking states 
\begin{equation}
\label{def:rb}
R_b =  \int_{1}^{E_{max}} p(E) dE ,
\end{equation}
which is zero for $\Lambda \leq 1$ and becomes positive for $\Lambda >1$.
$R_b$ is plotted in Figure \ref{figrb}b which shows that, for a given
value of $\eta/\alpha$, there exists a critical value for the nonlinear
coefficient $\Lambda$ such that the proportion of time  spent by the system 
in breaking states is maximal.
We can examine theoretically several particular cases.

\subsection{Absence of damping}
\begin{figure}
\begin{center}
\begin{tabular}{c}
\includegraphics[width=7.cm]{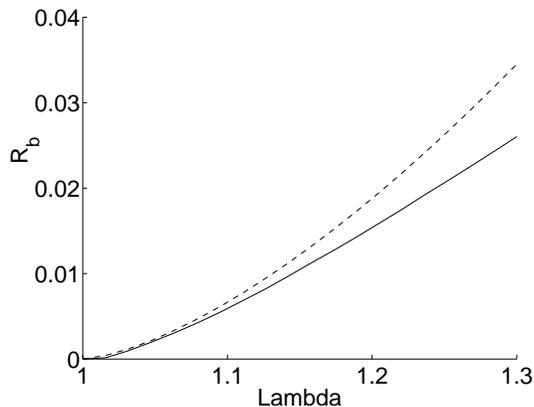}
\end{tabular}
\end{center}
\caption{
Proportion of time spent by the system 
in breaking states $R_b$ as a function of $\Lambda$.
The numerical integration  of Eq.~(\ref{def:rb})
is plotted in solid line, the approximate formula $R_b \simeq 0.21
\lambda^{3/2}$ is plotted in dashed lines.
\label{figrb0}
}
\end{figure}

In absence of damping $\eta=0$, the energy distribution does not
depend on the diffusion rate and it is given by
\begin{equation}
  p(E) = \frac{{\cal T}(E)}{\int_{- 1}^{E_{max}} {\cal T}(x) dx} .
\end{equation}
Let us first address the case where the number of atoms is just 
above the critical number that allows the existence of 
symmetry-breaking states.
If $\Lambda =1+\lambda$, $0 < \lambda \ll 1$,
then we get by using the expansions described 
in Appendix \ref{subsec:expand4} that
\begin{equation}
  p(E) \simeq \left\{
\begin{array}{ll}
\frac{1}{c [(1-E)/2]^{1/4}} 
K
\left( \frac{1}{2} -\frac{\sqrt{1-E}}{2\sqrt{2}}  \right) , \ \ \
&\mbox{ if } E \in [-1,1) ,\\
\frac{1}
{c [  \lambda + \sqrt{2 (E_{max}-E)} ]^{1/2} } 
K \left( \frac{2 \sqrt{2(E_{max}-E)}}{\lambda +\sqrt{2(E_{max}-E)}}  \right)
, \ \ \ 
&\mbox{ if } E \in (1,E_{max}],
\end{array}
\right.
\end{equation}
with $c  
\simeq 4.44$
and the proportion of time spent by the system 
in breaking states is $R_b \simeq 
0.21 {\lambda}^{3/2}$.
A comparison with the numerical integration of (\ref{def:rb})
shows that the approximate formula $R_b \simeq 0.21 \lambda^{3/2}$
is valid for $0 \leq  \lambda \leq 0.2$ (see Figure \ref{figrb0}).
Thus, for a small number of atoms, it is not likely to observe 
the MQL regime.

Let us now consider the case of a large number of atoms.
If $\Lambda \gg 1$, then we get by using the expansions described 
in Appendix \ref{subsec:expand3} that
\begin{equation}
  p(E) \simeq \left\{
\begin{array}{ll}
\frac{2}{\pi \sqrt{\Lambda}} K( \frac{1+E}{2} ) ,
&\mbox{ if } E \in [-1,1) ,\\
\frac{1}{\sqrt{2 \Lambda (E-1)}},
&\mbox{ if } E \in (1,E_{max}] ,
\end{array}
\right.
\end{equation}
and the proportion of time spent by the system 
in breaking states is $R_b \simeq 1-  2.55 \Lambda^{-1/2}$.
We can see in Figure \ref{figrb}b that this 
approximate formula is valid for $\Lambda \geq 40$.
Thus $R_b$ is close to one which means that it is very likely
to observe the MQT regime. 
We can also compute the transition time from one breaking state
to the other one.
This computation is based on a well-known result of stochastic
analysis which claims that  $\mu_{E_0,E_1} (E)$,
the mean time to exit the interval
$[E_0,E_1]$ starting from the energy $E$, satisfies ${\cal L}_E \mu =-1$
with the boundary conditions $\mu(E_0)=\mu(E_1)=0$.
The result is that the transition time is of order $\Lambda^2/\alpha$.
The computation of the transition time is similar to the one performed to
obtain the solution to the Kramers' exit problem which is
concerned with noise activated escape from a potential well \cite{hanggi}.
As a result we can predict that, for large $\Lambda$, the system
randomly goes from one breaking state to the other one according to 
a Markovian dynamics with an average
period of $\Lambda^2/\alpha$.

\subsection{Small damping}
In presence of damping Eq.~(\ref{eq:statE})
shows that the stationary energy distribution 
is shifted toward the low energy region.
To illustrate this phenomenon, we can consider and analyze 
the case $\Lambda \gg 1$ and $\eta/\alpha \ll 1/\Lambda$. 
Using the expansions described 
in Appendix \ref{subsec:expand3}, the energy distribution 
can be written as
\begin{equation}
  p(E) \simeq \left\{
\begin{array}{ll}
\frac{2C_\eta}{\pi \sqrt{\Lambda}} K( \frac{1+E}{2} ) ,
&\mbox{ if } E \in [-1,1) ,\\
\frac{C_\eta}{\sqrt{2 \Lambda (E-1)}} \exp 
\left( -2 \frac{\Lambda \eta}{\alpha} E^2 \right),
&\mbox{ if } E \in (1,E_{max}],
\end{array}
\right.
\end{equation}
with
\begin{equation}
  C_\eta = \left[ \int_0^{1/2} 
\frac{1}{\sqrt{2x}} 
\exp \left( -\frac{\Lambda^3 \eta}{\alpha} x^2 \right) dx
\right]^{-1} \simeq \left\{
\begin{array}{ll}
1
&\mbox{ if }  \eta/\alpha  \ll 1/\Lambda^3 \\
\frac{2 \sqrt{2}}{ \Gamma(1/4) } 
\frac{\eta^{1/4} \Lambda^{3/4}}{\alpha^{1/4} } 
&\mbox{ if } 1/ \Lambda^3 \ll \eta/\alpha  \ll 1/\Lambda
\end{array}
\right.
\end{equation}
As a result, the proportion of time spent by the system 
in breaking states is $R_b \simeq 1-  2.55 C_\eta {\Lambda}^{-1/2}$
which decays with $\eta$.
It is less and less likely to observe the MQT regime
as the damping becomes larger.

\subsection{Strong damping}
Assume that $ \eta / \alpha \gg 1/(1+\Lambda)$.
Using the expansions derived in Subsection \ref{subsec:expand1}
 the diffusion operator can be written as
\begin{equation}
  {\cal L}_E  = \alpha \frac{\partial}{\partial E}
(1+E)  \frac{\partial}{\partial E} - \eta 
(1+\Lambda)(1+E)  \frac{\partial}{\partial E} .
\end{equation}
The stationary energy distribution is
\begin{equation}
  p(E) = \frac{1}{E_c} \exp \left( - \frac {1+E}{E_c} \right)
{\bf 1}_{[-1,E_{max}]}(E) ,
\end{equation}
where $E_c= \alpha /[ (1+\Lambda) \eta]$.
The average energy is $-1+E_c$.
The proportion of time spent in breaking states is exponentially small
$R_b \simeq \exp(-2 / E_c)$.

\subsection{Numerical experiments}
In this section we compare our theoretical predictions
with numerical simulations of Eqs.~(\ref{eq:sys1a}-\ref{eq:sys1b}).
The random fluctuations of $\Delta E(t)$
are modeled by a stepwise constant process 
$$
\Delta E(t) = \sigma \sum_{j} X_j {\bf 1}_{[ j t_c, (j+1)t_c)}(t) ,
$$
where the $X_j$ are independent and identically distributed 
random variables with uniform distribution  over $(- \sqrt{3}
, \sqrt{3} )$ and $t_c$ is the coherence time.
The coefficient $\alpha$ is then given by
$$
\alpha = \frac{\sigma^2 t_c}{2} .
$$
Damping is absent $\eta=0$ and
the parameter $\Lambda=15$.
Three series of simulations are performed with the parameters
$(\sigma,t_c)=(4,0.05)$, $(\sigma,t_c)=(3,0.1)$,
and $(\sigma,t_c)=(2,0.05)$.
Note that the first two configurations give almost the same value
for the effective parameter $\alpha \simeq 0.4$.
We have carried $1000$ simulations for each configuration.
The initial conditions are $z_0=0$, $\phi_0=0$,
but we  have checked that the initial conditions play no role 
in the long-time dynamics.

\begin{figure}
\begin{center}
\begin{tabular}{cc}
{\bf a)}
\includegraphics[width=7.cm]{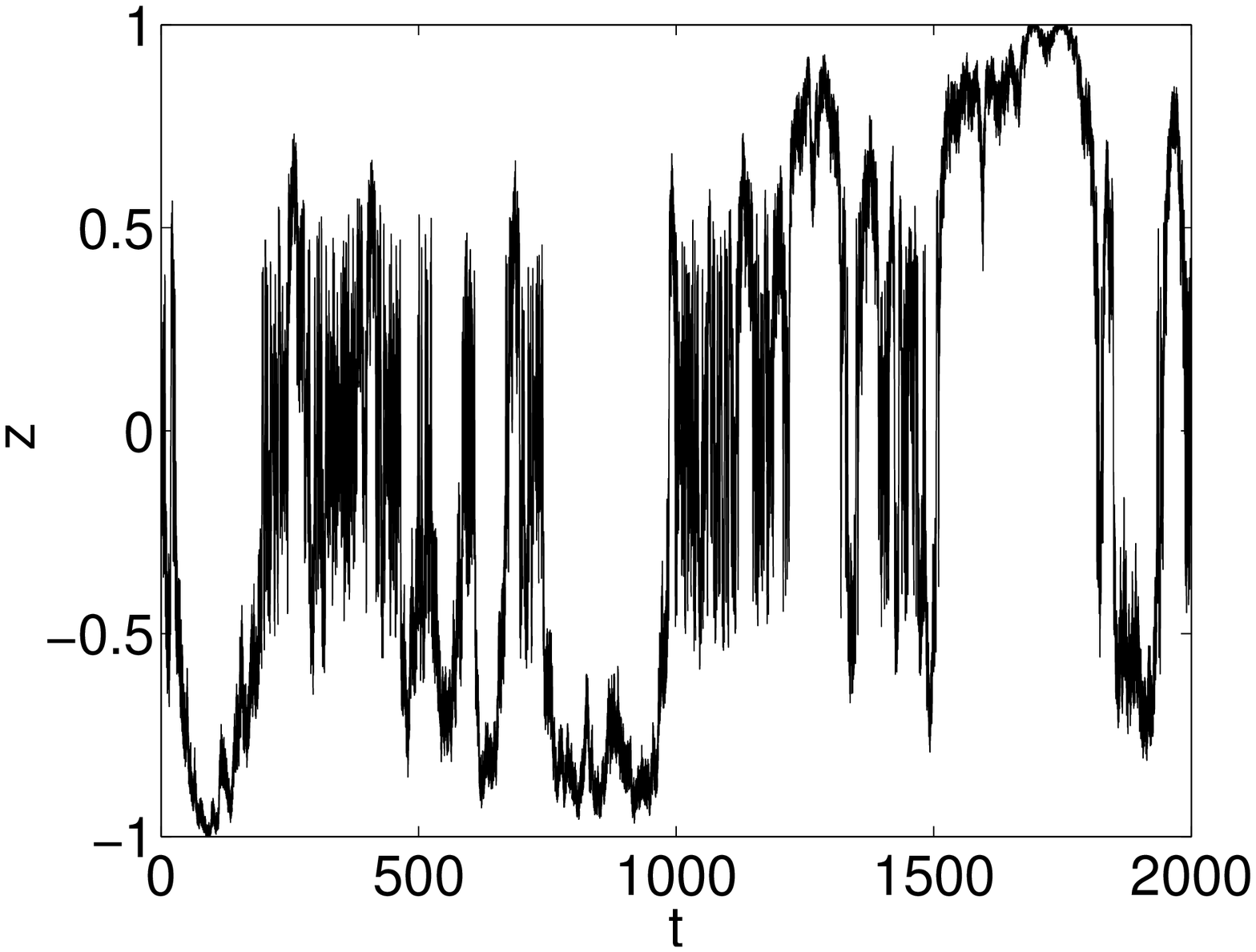}
&
{\bf b)}
\includegraphics[width=7.cm]{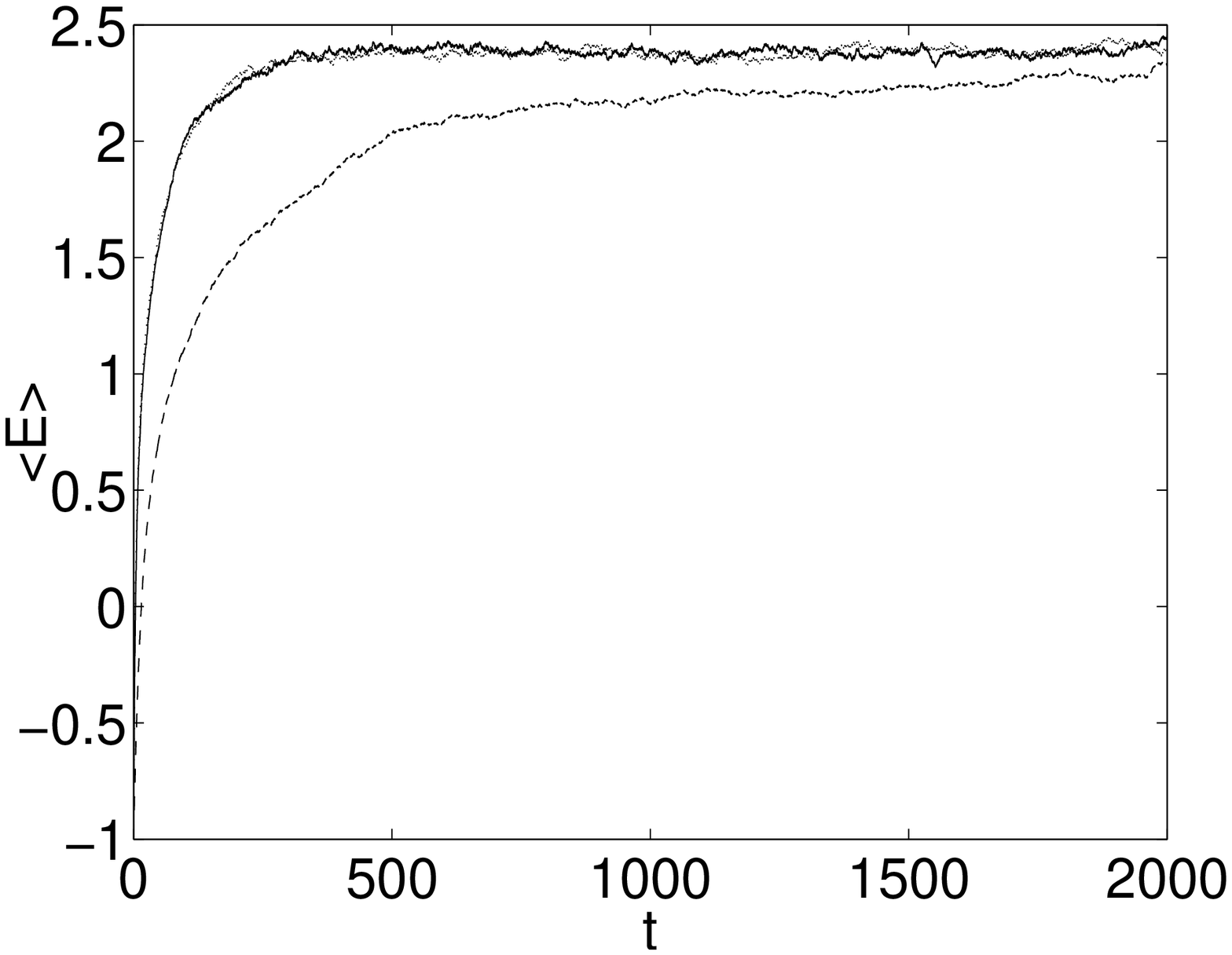}
\end{tabular}
\end{center}
\caption{Picture a:
A typical $z$-trajectory for the set of parameters
$(\sigma,t_c)=(4,0.05)$.
Picture b:
Average energy versus time for 
three different configurations:
$(\sigma,t_c)=(4,0.05)$ (solid line), $(\sigma,t_c)=(3,0.1)$ (dotted line),
and $(\sigma,t_c)=(2,0.05)$ (dashed line).
\label{figsim}
}
\end{figure}

In Figure \ref{figsim}a we plot a typical $z$-trajectory for the
configuration $(\sigma,t_c)=(4,0.05)$.
As predicted by the theory, noise induces an energy diffusion
which makes the system visit symmetry-breaking states.
More quantitatively, 
the system spends around half-time is symmetry-breaking states
as predicted by Figure \ref{figrb}b,  and the 
life-time  of a symmetry-breaking state is of the order 
of $\Lambda^2/\alpha \simeq$ a few hundreds.

In figure \ref{figsim}b we plot the energy of the system
for the three configurations. The energy is averaged 
over the $1000$ simulations.
We can observe that the behavior is the same for the 
first two configurations,
which is in agreement with the fact that both configurations
possess the same effective parameter $\alpha \simeq 0.4$.
We can also compare the asymptotic value of the average energy
with the theoretical one $\left<E \right> \simeq \Lambda /6 =2.5$.
Note that the time necessary to reach the asymptotic value is 
of the order of $\Lambda^2/\alpha \simeq 500$.
The  configuration $(\sigma,t_c)=(2,0.05)$ corresponds 
to a smaller value of $\alpha = 0.1$.
As a result the time necessary to reach the asymptotic energy value is
longer,
but the asymptotic value itself is the same as for the other
configurations as it depends only on $\Lambda$.

\section{Conclusion}
In this paper we have considered the dynamics of two weakly-coupled 
BECs.
We have investigated the influence of 
small random oscillations of the barrier-laser position which induce
fluctuations of the zero-point energies of the double-well trap.
We have shown that random noise induces an energy diffusion which 
allows the system to visit symmetry-breaking states when the number of
atoms (proportional to $\Lambda$) exceeds a threshold value.
We have shown that 
the energy distribution evolves to a stationary distribution
which depends on the initial state of the BEC only through the total
number of atoms, and not through the initial imbalance or phase
difference.
We have computed the time necessary for the loss of memory of the
initial conditions and the establishment of the stationary dynamics.
Then we have described the stationary dynamics of the BEC
which visits symmetric and symmetry-breaking states according to a
random Markovian dynamics.
We have investigated the role of damping in this process.
In particular we have shown that, for a given damping, there
exists a critical value for the number of atoms where
the proportion of time spent by the BEC in symmetry-breaking states
(i.e. in a macroscopic quantum localization state)
is maximal.

\appendix

\section{Expansions}

\subsection{$E$ close to $E_{min}$}
\label{subsec:expand1}%
The minimal energy is $E_{min}=-1$.
We can expand all quantities when $E=E_{min}+\de$, $\de \geq 0$.
We get in particular 
$M \simeq \frac{\Lambda^2}{2(1+\Lambda)^2} \de $,
${z_2} \simeq \frac{\sqrt{2}}{\sqrt{1+\Lambda}}
\sqrt{\de}$,
\begin{eqnarray}
   && {\cal T}(E) = \frac{2 \pi}{\sqrt{1+\Lambda}} ,\\  
   && {\cal Z}(E,\theta) =   \frac{\sqrt{2}}{\sqrt{1+\Lambda}}
\sqrt{\de} \cos(\theta) ,\\
   && {\cal C}(E,\theta) =   1-\de \sin(\theta)^2 ,
\end{eqnarray}
so that
\begin{eqnarray}
  &&{\cal N}(E) = (1+\Lambda) \de ,\\
  &&{\cal B}(E) = \frac{2 \de}{1+\Lambda} .
\end{eqnarray}

\subsection{$E$ close to $E_{max}$}
\label{subsec:expand2}%
Here we assume that $\Lambda >1$ so that $E_{max}>1$.
We can expand all quantities when $E=E_{max}-\de$, $\de \geq 0$.
We get in particular 
$M \simeq \frac{\sqrt{2} (\Lambda^2-1) }{8 \sqrt{ \Lambda\de}} $,
${z_2} \simeq \frac{\sqrt{\Lambda^2-1}}{\Lambda}+
 \frac{\sqrt{2 \de}}{\sqrt{\Lambda(\Lambda^2-1)}}$,
\begin{eqnarray}
   && {\cal T}(E) = \frac{2 \pi}{\sqrt{\Lambda^2-1}} ,\\  
   && {\cal Z}(E,\theta) =   \frac{\sqrt{\Lambda^2-1}}{\Lambda}+
\frac{ \sqrt{2\de} }{\sqrt{\Lambda(\Lambda^2-1)}} \cos(\theta) ,\\
   && {\cal C}(E,\theta) =   -1+ \Lambda \de \sin^2(\theta),
\end{eqnarray}
so that
\begin{eqnarray}
  &&{\cal N}(E) = \Lambda  (\Lambda^2-1) \de ,\\
  &&{\cal B}(E) = \frac{ \de}{ \Lambda  (\Lambda^2-1)}.
\end{eqnarray}

\subsection{Large $\Lambda$}
\label{subsec:expand3}%
Let us assume that $\Lambda \gg 1$.
If $E \in [-1,1)$, then $M\simeq (1+E)/2<1$,
\begin{eqnarray}
   && {\cal T}(E) = \frac{4}{\sqrt{\Lambda}} K
\left( \frac{1+E}{2} \right) , \\  
   && {\cal Z}(E,\theta) =   \frac{\sqrt{2}}{\sqrt{\Lambda}}
\sqrt{1+E} {\rm cn} \left( \frac{2 K(\frac{1+E}{2}) \theta}{\pi},
  \frac{1+E}{2} \right)  .
\end{eqnarray}

If $E\in ( 1 , E_{max}]$, then by setting $E = \Lambda e$, $e \in (0,1/2]$,
we have
\begin{eqnarray}
   && {\cal T}(E) = \frac{\sqrt{2} \pi}{\Lambda \sqrt{e}} , \\  
   && {\cal Z}(E,\theta) =   \sqrt{2 e} \left[ 1 -\frac{\sqrt{1-2e}}{\Lambda e}
\sin^2(\frac{\theta}{2}) \right] ,
\end{eqnarray}
so that
\begin{eqnarray}
  &&{\cal N}(E) =2 \Lambda^2 e, \\
  &&{\cal B}(E) = \frac{1-2e}{2 \Lambda^2 e}.
\end{eqnarray}

\subsection{$\Lambda$ just above $1$}
\label{subsec:expand4}%
This case is interesting in that it is the regime where
symmetry-breaking appear.
Let us assume $\Lambda=1+\lambda$, $0 < \lambda \ll 1$.
If $E \in[-1,1)$, then $M \simeq \frac{1}{2} ( 1-
\frac{\sqrt{1-E}}{\sqrt{2}})$ and
$$
{\cal T}(E) = \frac{2 \sqrt{2}}
{ {E'}^{1/4}} 
K
\left( \frac{1}{2} (1-\sqrt{E'})  \right) , \ \ \ \ \
E'=(1-E)/2 .
$$
Note that $E_{max}  = 1+\lambda^2/2+O(\lambda^3)$.
If $E=1 +\frac{\lambda^2}{2}e$, $e \in [0,1]$, then
$M \simeq \frac{1}{2} ( 1+ 
\frac{1}{\sqrt{1-e}})$ and
$$
{\cal T}(E) = \frac{2 \sqrt{2}}
{\sqrt{  1+ \sqrt{e'} } \sqrt{\lambda}} 
K \left( \frac{2 \sqrt{e'}}{1+\sqrt{e'}}  \right)  , \ \ \ \ \
e'=1-e.
$$

\end{document}